\documentclass{article}
\usepackage{spconf,amsmath,graphicx}
\usepackage{multirow}
\usepackage{amsxtra}
\usepackage{amsmath}
\usepackage{threeparttable,comment,bm}
\usepackage{cite,array,amssymb}
\usepackage{subfig}
\usepackage{mathrsfs}
\usepackage{mathtools}
\usepackage{color}

\usepackage{fancyhdr}
\newenvironment{itemize_small}
{
\begin{list}{$\bullet$\ \ }
  {\setlength{\itemindent}{0pt}
  \setlength{\leftmargin}{1cm}
  \setlength{\labelsep}{0.2cm}
  \setlength{\itemsep}{0.5em}
  \setlength{\parsep}{0em}
  }
  }{\end{list}
}
\begin{document}\sloppy
\makeatletter

\renewcommand\section{\@startsection{section}{1}{\z@}
{0.5ex \@plus 0ex \@minus -2ex}
{0.5ex \@plus0ex}
{\normalfont\Large\bfseries}}
\renewcommand\subsection{\@startsection{subsection}{2}{\z@}
{0.5ex \@plus 0ex \@minus -2ex}
{0.5ex \@plus0ex}
{\normalfont\Large\bfseries}}
\makeatother
\title{grayscale-based block scrambling image encryption for social networking
services}

%\IEEEauthorblockA{\IEEEauthorrefmark{1}Tokyo Metropolitan
%University, Tokyo, 191-0065, Japan}
%\IEEEauthorblockA{\IEEEauthorrefmark{2}Chiba University,
%Chiba, 263-8522, Japan}
\name{Warit Sirichotedumrong$^*$, Tatsuya Chuman$^*$, Shoko
Imaizumi$^\dagger$ and Hitoshi Kiya$^*$
\thanks{This work was partially supported by Grant-in-Aid for Scientific Research(B), No.17H03267, from the Japan Society for the Promotion Science.}}
\address{$^*$Tokyo Metropolitan University, Asahigaoka, Hino-shi, Tokyo,
191-0065, Japan \\ $^\dagger$Chiba University, Chiba, 263-8522, Japan}

\maketitle
\begin{abstract}
This paper proposes a new block scrambling encryption scheme that enhances the
security of encryption-then-compression (EtC) systems for JPEG images, which are
used, for example, to securely transmit images through an untrusted channel provider. The proposed method allows the use of a smaller block size and a larger
number of blocks than the conventional ones. Moreover, images encrypted using
proposed scheme include less color information due to the use of grayscale even when the
original image has three color channels. These features enhance security against
various attacks such as jigsaw puzzle solver and brute-force attacks. The
results of an experiment in which encrypted images were uploaded to and then
downloaded from Twitter and Facebook demonstrated the effectiveness of the
proposed scheme for EtC systems.
\end{abstract}
\begin{keywords}
Encryption, EtC system, social media
\end{keywords}
\section{Introduction}
\label{sec:intro}
The rapid growth of the Internet and multimedia systems causes the increment of using images and video especially in Social Networking Service (SNS). 
To securely transmit images through an untrusted channel provider such as SNS,
encryption has to be performed.
There are many encryption schemes which have been studied for securing an
image\cite{huang2014survey,lagendijk2013encrypted,zhou2014designing}.
The most secure option is to fully encrypt the whole image using the famous
cryptosystems, such as RSA and AES. However, security is not the only requirement that cryptosystems should satisfy. 
Low processing cost and the format compliance have to be considered for using in many applications. 
A lot of perceptual encryption schemes has been proposed to satisfy the requirements by trading-off the security of encryption schemes\cite{Zeng_2003,Ito_2008,Kiya_2008,Ito_2009,Tang_2014,Chengqing}.\par 
This paper focuses on encrypting images before compression process which is called Encryption-then-Compression (EtC) systems\cite{zhou2014designing,Erkin_2007,nimbokar2014survey}. 
For transmitting the encrypted images over the internet via many platforms, 
the format of encrypted image has to be compatible with the international image
compression standards.
The previously proposed EtC
systems\cite{Johnson_2004,Liu_2010,Zhang_2010,Hu_2014,zhou2014designing} is not
compatible with with the international standards.
Consequently, the encryption schemes for EtC systems with format compliance to
international standards have been proposed\cite{watanabe2015encryption,kurihara2015encryption,KURIHARA2015,KuriharaBMSB,Kuri_2017}.
Furthermore, the security of conventional EtC systems against jigsaw puzzle and
brute-force attacks have been discussed and
evaluated\cite{CHUMAN2017ICASSP,CHUMAN2017ICME}.\par Considering the
applicability of EtC systems to SNS, it has been confirmed that the conventional
EtC is applicable to SNS. However, color components of images with
the conventional scheme can be affected by JPEG compression.
Due to the limitation, the color image must be split into $16 \times 16$-blocks.
Because of such a situation, this paper proposes a new encryption scheme to
improve format compliance, robustness against SNS recompression, and the
security against several attacks. The contributions of this work are:
\vspace{-0.3cm}
\begin{itemize_small}
	\item We propose a new block scrambling encryption scheme for EtC systems which
	enhances the security by dividing the image into smaller block size, having a
	large number of blocks, and containing less color information.
	\item We discuss the important characteristics of the image recompression
	manipulated by SNS provider and elaborately consider this characteristic for
	designing the encryption scheme which is appropriate for SNS.
	\item We evaluate the security and downloaded image quality of
	the proposed scheme and compare them with the conventional EtC
	encryption scheme.
\end{itemize_small}
\vspace{-0.3cm}
\par
The rest of paper is organized as follows. Section\,\ref{etc} introduces the EtC
systems for image encryption which describe block scrambling-based encryption
procedures, the security of EtC systems, and application to SNS.
Section\,\ref{grayenc} elaborates the  grayscale-based block scrambling
encryption scheme, starting from the grayscale encryption processes followed by
describing how the proposed scheme enhances the security and avoids image
manipulation process by SNS providers. We evaluate and discuss the performance
in Section\,\ref{exp}. Concluding remarks are in Section\,\ref{conclusion}.
\section{Preparation}
\label{etc}
In this section, after the conventional block scrambling-based image
encryption\cite{kurihara2015encryption,KURIHARA2015,KuriharaBMSB,Kuri_2017} is
summarized, the security of the scheme against ciphertext-only attacks is
addressed. Then, the application of EtC systems for SNS is discussed.
\begin{figure}[t]
\centering
\includegraphics[width =8.6cm]{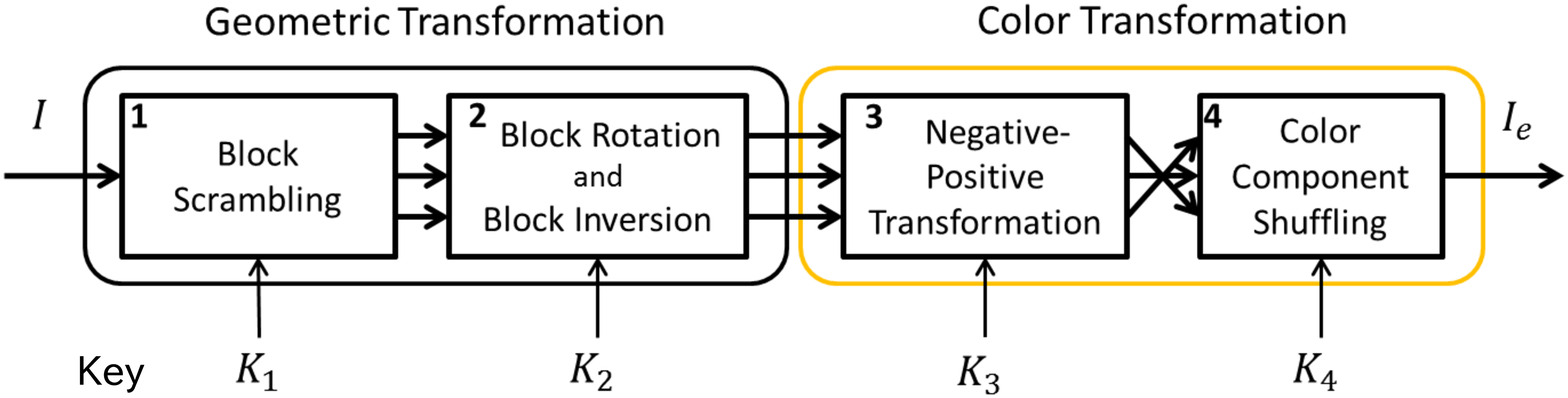}
\caption{Block scrambling-based processing
steps{\cite{CHUMAN2017ICASSP,CHUMAN2017ICME}}}
\label{fig:blockbased}
\end{figure}
\begin{figure}[!t]
\captionsetup[subfigure]{}
\centering
\subfloat[Original image\newline($X \times Y$ = $672 \times 480$)]{\includegraphics[clip, height=2.5cm]{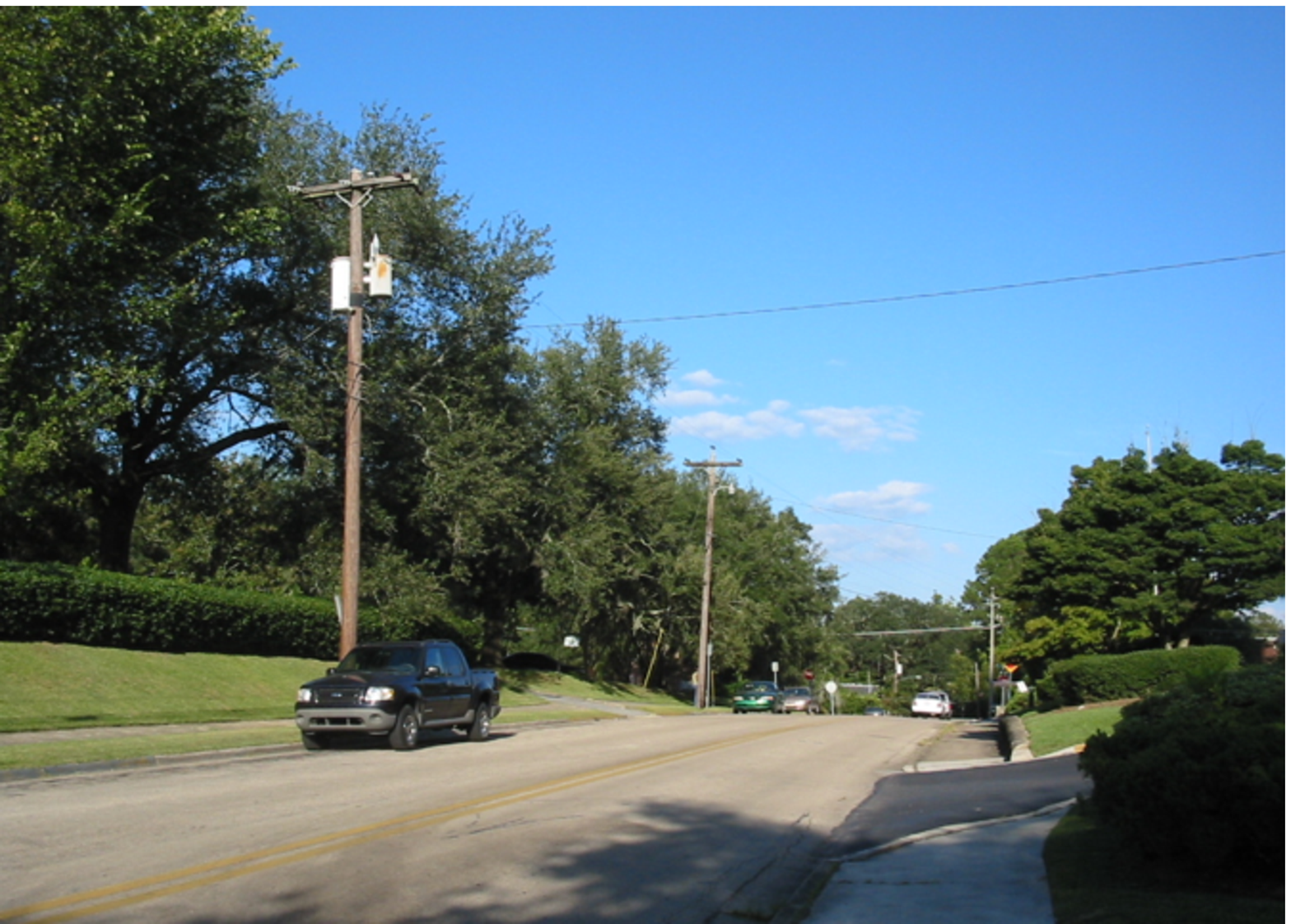}
\label{fig:label-B}}
\hfil
\vspace{-2mm}
\subfloat[Encrypted image\cite{kurihara2015encryption,KURIHARA2015}\newline($B_{x}\!=\!B_{y}\!=\!16, n\!=\!1260$)]{\includegraphics[clip, height=2.5cm]{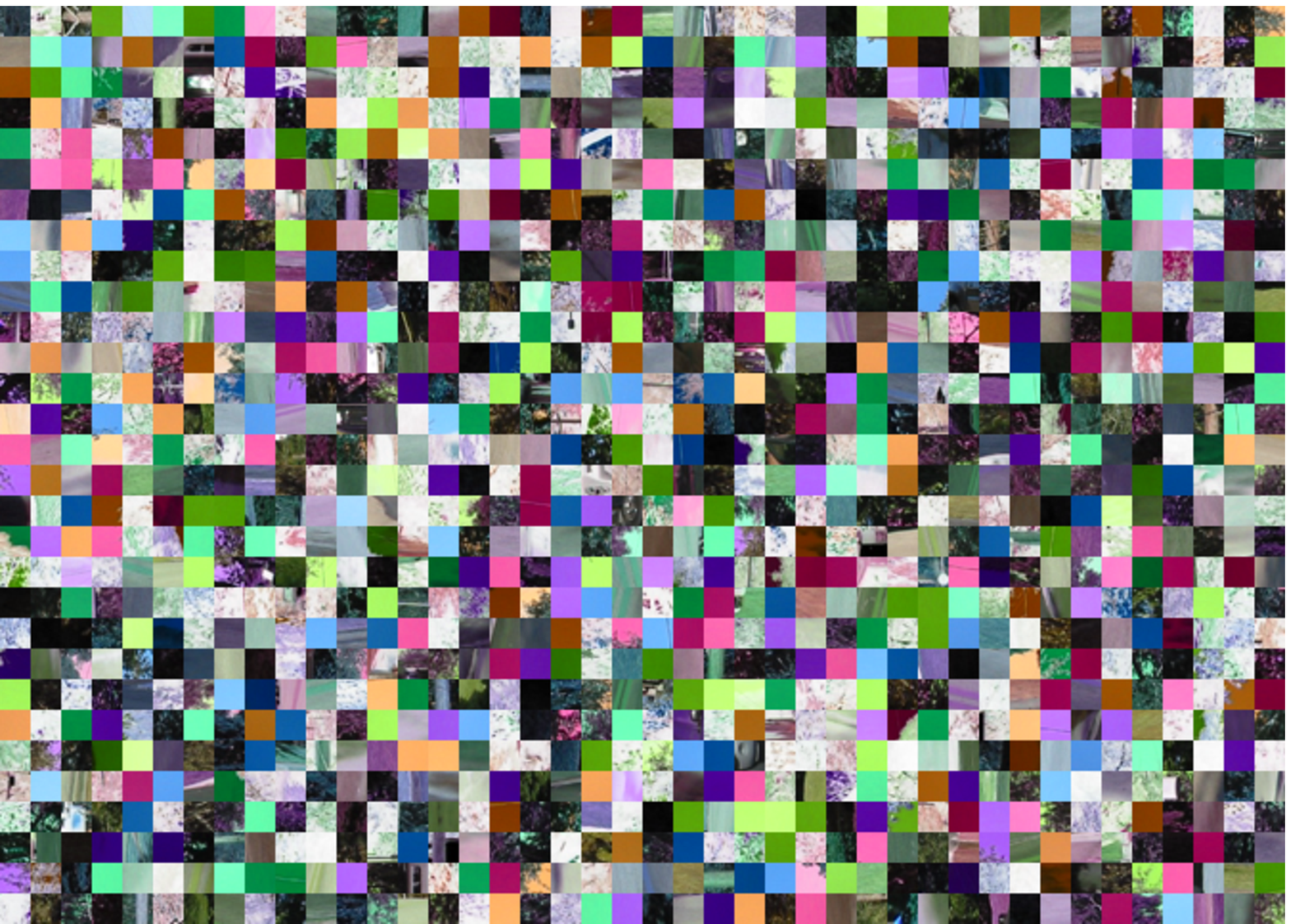}
\label{fig:label-C}}
\\
\subfloat[Encrypted image using proposed scheme ($B_{x}=B_{y}=8, n=15120$)]{\includegraphics[clip, height=2cm]{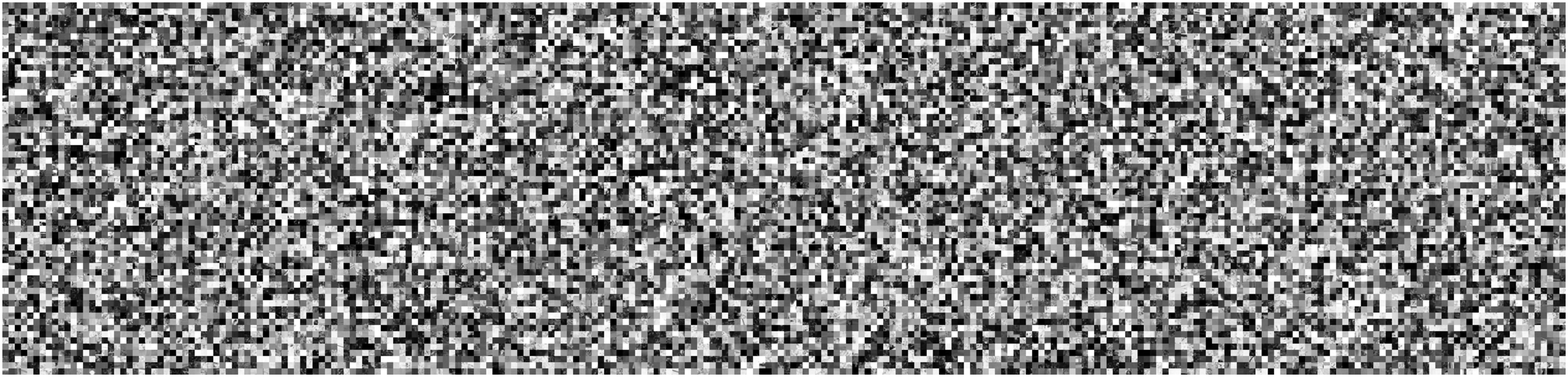}}
\caption{Examples of encrypted images}
\label{fig:eximages}
\vspace{-0.5cm}
\end{figure}
\subsection{Block scrambling-based image encryption}
\label{blocksc}
%In order to encrypt \textcolor{red}{an} image using EtC system\cite{kurihara2015encryption,KURIHARA2015,KuriharaBMSB,Kuri_2017}, an image $I$ is encrypted to the encrypted image called $I_e$ by performing block scrambling-based image encryption scheme which is describe as the following five steps.
According to the block scrambling-based image encryption scheme for EtC
systems\cite{watanabe2015encryption,kurihara2015encryption,KURIHARA2015,KuriharaBMSB,Kuri_2017},
an image with $M \times N$ pixels is divided into non-overlapping blocks each
with $B_x \times B_y$ pixels. The number of divided blocks, $n$, is expressed by
\begin{equation}
\label{eq:blocknum}
n = \lfloor \frac{M}{B_x} \rfloor \times \lfloor \frac{N}{B_y} \rfloor
\end{equation}
where $\lfloor \cdot \rfloor$ is the round down function to the nearest
integer.\par
In order to generate an encrypted image ($I_e$), each divided block is processed
using the following four block scrambling-based steps (See
Fig.\,\ref{fig:blockbased}).
\begin{itemize_small}

\item[Step1:] Divide an image with $M \times N$ pixels ($I$) into blocks with
$B_x \times B_y$ pixels, and permute the divided blocks randomly based on the random integer which is generated by a secret key $K_1$.

\item[Step2:] Randomize the integer using a secret key $K_2$, then rotate and
invert each block according to the previously randomized integer.

\item[Step3:] Apply the negative-positive transformation to each block using a
random binary integer generated by a secret key $K_3$. A transformed pixel of
$i$th block is represented by $p'$ and computed as
\begin{equation}
p'=
\left\{
\begin{array}{ll}
p & (r(i)=0) \\
p \oplus (2^L-1) & (r(i)=1)
\end{array}
\right.
\end{equation}
where $r(i)$ is a random binary integer generated by $K_3$ and $p$ is the pixel value of an original image with $L$ bits per pixel.

\item[Step4:] The three color components in each block are shuffled using a
senary integer generated by the fourth secret key $K_4$.
\end{itemize_small}
Note that the secret keys, $K_1$, $K_2$, and $K_3$, are commonly used for all color components.\par
According to the block scrambling-based encryption scheme, the above four
steps construct $I_e$ that provides a compatibility with the JPEG standard and
almost preserves the compression efficiency at the same level as the original JPEG image.
In JPEG compression, color sub-sampling is usually done for reducing the color
components of a color image. In order to make 8 $\times$ 8-blocks for
color sub-sampling process, the color image must be split into Minimum Coded
Unit (MCU) which corresponds to 16 $\times$ 16-blocks. Therefore, the possible
smallest block size of block scrambling-based encryption is 16 $\times$ 16.
When the block size is smaller than 16 $\times$ 16, the block scrambling
process affects the color sub-sampling of JPEG compression.
An example of an encrypted image ($B_{x}=B_{y}=16$) is illustrated in Fig.\,\ref{fig:eximages}(b) where Fig.\,\ref{fig:eximages}(a) is the original one.
\subsection{Security analysis}
\label{blocksec}
Along with the encryption scheme, robustness and security against the attacks
have to be considered in terms of key space as explained in
\cite{KURIHARA2015}.
The key space of the scheme is generally large enough against the brute-force attacks as ciphertext-only attack.
However, regarding the blocks of an encrypted image as pieces of a jigsaw
puzzle, jigsaw puzzle solver attacks based on the correlation can be
assumed\cite{Son_2014_ECCV,Paikin_2015_CVPR,Sholomon_2016_GPEM,Son_2016_CVPR,Sholomon_2014_AAAI,Andalo_2016}.
For example, the jigsaw puzzle solver\cite{Sholomon_2016_GPEM} completely succeeded in assembling large puzzles
which consist of 30745 pieces with the size of $28 \times 28$. Besides, even
when the number of blocks in an encrypted image is larger than 22755, there is a
possibility that the image is completely decrypted if the piece size is
large\cite{Paikin_2015_CVPR}.
There are three conditions making assembling encrypted images more difficult as indicated below.
\begin{itemize_small}
	\item The number of blocks is large.
	\item Block size is small.
	\item Encrypted images include JPEG distortion\cite{CHUMAN2017IWSDA}.
\end{itemize_small}
Moreover, as most of jigsaw puzzle solvers utilize the color information to
assemble the encrypted images, reducing the color components makes the assembling process
more difficult.
\begin{figure}[t]
\centering
\includegraphics[width =7.6cm]{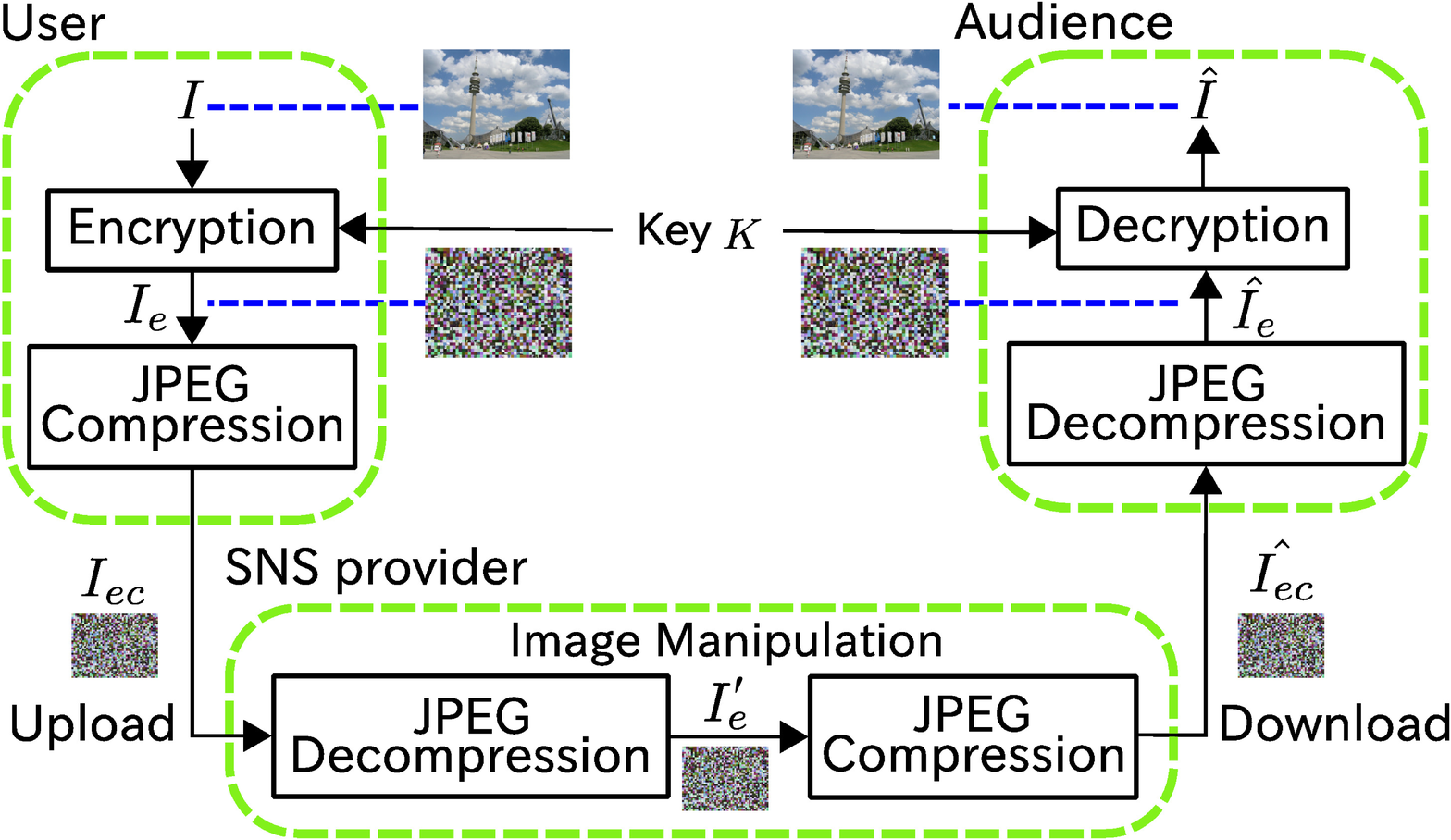}
\caption{EtC system{\cite{CHUMAN2017ICASSP,CHUMAN2017ICME}}}
\label{fig:etc}
\vspace{-0.5cm}
\end{figure}
\begin{table*}
\centering
\caption{Relationship between uploaded JPEG files and downloaded ones in terms
of sub-sampling ratios\cite{CHUMAN2017APSIPA}}
\vspace{-0.3cm}
\begin{threeparttable}
\scalebox{1}{
\begin{tabular}{|c||c|c||c|c|}
\hline
\multirow{2}{*}{SNS provider} & \multicolumn{2}{c||}{Uploaded JPEG file} & \multicolumn{2}{c|}{Downloaded JPEG file} \\ \cline{2-5} 
 & \begin{tabular}[c]{@{}c@{}}Sub-sampling ratio\end{tabular} & $Q_{f}$ & \begin{tabular}[c]{@{}c@{}}Sub-sampling  ratio\end{tabular} & $Q_{f}$ \\ \hline
\multirow{4}{*}{Twitter (Up to 4096$\times$4096 pixels)} & \multirow{2}{*}{4:4:4} & low & \multicolumn{2}{c|}{No recompression} \\ \cline{3-5} 
 &  & high & 4:2:0 & 85 \\ \cline{2-5} 
 & \multirow{2}{*}{4:2:0} & 1,2,\ldots84 & \multicolumn{2}{c|}{No recompression} \\ \cline{3-5} 
 &  & 85,86,\ldots100 & 4:2:0 & 85 \\ \hline
\multirow{2}{*}{\begin{tabular}[c]{@{}c@{}}Facebook (HQ, Up to 2048$\times$2048
pixels)\\ Facebook (LQ, Up to 960$\times$960 pixels)\end{tabular}} & 4:4:4 & \multirow{2}{*}{1,2,\ldots100} & \multirow{2}{*}{4:2:0} & \multirow{2}{*}{71,72,\ldots85} \\ \cline{2-2} & 4:2:0 &  &  &  \\ \cline{1-5} 
\end{tabular}
}
\end{threeparttable}
\label{tb:sns_manipulation}
\vspace{-0.2cm}
\end{table*}
\subsection{Application to SNS}
\label{apptosns}
It has been confirmed that EtC systems can be applied to SNS\cite{CHUMAN2017APSIPA}.
Figure\,\ref{fig:etc} illustrates the application of EtC systems for SNS,
 where a user wants to securely transmit
 image $I$ to an audience, via a SNS provider.
 As the user does not give the secret key $K$ to the SNS provider,
 the privacy of shared image is controlled by the user,
 even if the SNS provider decompresses image $I$.
 Therefore, the user is able to protect the privacy by him/herself. 
 Although encrypted images saved in the SNS servers are leaked by malicious
 users, the third party and general audiences could not see these images
 visually unless they have the secret keys.
\par Meanwhile, it is known that almost all SNS providers manipulate images
uploaded by users, e.g., rescaling image resolution and recompressing with different
parameters, for decreasing the data size of images\cite{giudice2016_arxive,Moltisanti2015_ICIAP,CHUMAN2017APSIPA}.
As a result, the quality of images recompressed by SNS providers is reduced by image manipulation on social media.
\par This paper proposes a new encryption scheme for the EtC system, which
avoids some effects of recompression by SNS providers.
As a result, using the proposed scheme enables us to secure the privacy of the
image and keep higher quality than the conventional one, even when
recompression is carried out by SNS providers.
\section{Proposed scheme}
\label{grayenc}
In this section, the proposed block scrambling-based image encryption scheme is
described, and then the security of the scheme is discussed.
\subsection{Encryption procedure}
\label{proposed}
Three steps to encrypt a color image with $M \times N$ pixels are carried
out as follows.
\begin{itemize_small}
	
\item[Step1:] The RGB color components of the full-color image are separated into three individual channels. 
The scheme considers each channel as an individual image, and red, green, and
blue channels can be respectively represented by $i_r$, $i_g$, and $i_b$.

\item[Step2:] $i_r$, $i_g$, and $i_b$ are combined as a new image in
grayscale ($I_{gray}$). For example, this combination process can be done
vertically and horizontally, so size of the new image is equal to $M \times 3N$ or $3M \times N$.

\item[Step3:] Step 1 to step 3 of block scrambling-based encryption
described in Section\,\ref{blocksc} is performed over $I_{gray}$.
\end{itemize_small}\par
After three steps above, the grayscale encrypted image is produced and can be
represented by $I_{e_{gray}}$. Note that the color components shuffling in the
conventional scheme are neglected.
\subsection{Smaller block size with less chroma component}
\label{block_size}
Since the JPEG standard downsamples chroma components by splitting a color
image into MCU, the possible smallest block size of conventional scheme is
$B_{x} = B_{y} = 16$.
As a grayscale image contains only one color channel
per pixel, it is not sub-sampled by JPEG compression. Hence, the
smallest block size of the proposed scheme is decreased to $B_{x} = B_{y} = 8$
which causes fourfold rising to the number of blocks. As a result, as shown in
Fig.\,\ref{fig:eximages}(c), the size of encrypted image is $3(M \times
N)$, the number of block is totally 12 times larger than that with the
conventional one.
Summarily, the proposed scheme enhances the security by
utilizing less color information, block size reduction, and larger number of
blocks.
\subsection{Image manipulation on SNS}
\label{graysns}
The proposed scheme also has some advantages over the SNS platform. This
paper focuses on the JPEG compression as one of compression methods because the
JPEG standard is the most widely used image compression standards, and most of SNS providers support the JPEG standard\cite{CHUMAN2017APSIPA}. 
\par Generally, the JPEG standard encodes color images by transforming the color components from RGB space to YCbCr space, then the color components, Cb and Cr, are downsampled to reduce the spatial
resolution.
Besides, when an image is uploaded to SNS, the
image will be recompressed again by providers as shown in
Table\,\ref{tb:sns_manipulation}\cite{CHUMAN2017APSIPA}.
The image recompression of SNS providers is concluded as follows.
\begin{itemize}
\item {\bf Twitter} decides whether an uploaded image has to be manipulated or
not based on uploaded image compression properties as shown in
Table\,\ref{tb:sns_manipulation}. When the quality factor ($Q_f$) of  an
uploaded JPEG image is higher than 84, Twitter transcodes it into the new image.
$Q_f$ of transcoded image is equal to 85, and the color components of transcoded
image are distorted using 4:2:0 color sub-sampling.
Otherwise, Twitter does not recompress the uploaded image.
\item {\bf Facebook} always recompresses an uploaded images regardless of sub-sampling ratio and $Q_f$. 
Every image is recompressed with 4:2:0 color sub-sampling, and there are many possible $Q_f$ using in Facebook recompression process. 
Depending on Facebook compression algorithm\cite{CHUMAN2017APSIPA}, $Q_f$ is
selected from the specific range (71$\leqq$$Q_{f}$$\leqq$85).
\end{itemize}
\par
According to \cite{CHUMAN2017APSIPA}, it has been confirmed that the
conventional block scrambling scheme is also applicable to SNS while
other encryption schemes, such as AES, cannot be applied to SNS. Moreover, the
proposed scheme can reduce influence of the color sub-sampling, caused by SNS
providers.
\section{Experimental Results}
\label{exp}
\subsection{Experimental set-up}
\label{set-exp}
In order to evaluate some performances of the encryption schemes, we employed
two datasets as below.
\vspace{-0.17cm}
\begin{itemize}
\item[(a)] 20 images from MIT dataset ($672 \times
480$)\cite{Cho_2010_CVPR}.
\item[(b)] 20 images from resized Ultra-Eye dataset ($240 \times
128$)\cite{Nemoto_2014_MEX}.
\end{itemize}
\vspace{-0.17cm}
All images in both datasets were encrypted using the conventional scheme ($B_x =
B_y = 16$ and $B_x = B_y = 8$) and the proposed scheme ($B_x = B_y = 8$).
We also compressed the images using the JPEG standard from
Independent JPEG Group (IJG) software\cite{JPEGLIB} with specific range of quality factors, $Q_f \in [70,100]$.
We focused on Twitter and Facebook because these SNS providers always recompress
if images uploaded by users meet the conditions. The encrypted and non-encrypted
JPEG files with 4:4:4 sampling from dataset (a) were uploaded to
Twitter and Facebook while the encrypted and non-encrypted images from dataset
(b) were utilized for evaluating the robustness against
jigsaw puzzle solver
attacks\cite{Gallagher_2012_CVPR,Son_2016_CVPR,Cho_2010_CVPR,CHUMAN2017ICASSP,CHUMAN2017ICME}.
 \begin{figure}
 \captionsetup[subfigure]{justification=centering}
\centering
\begin{minipage}{\linewidth}
\centering
\subfloat[Original image\newline($X \times Y$ = $240 \times
128$)]{\includegraphics[height=2.2cm]{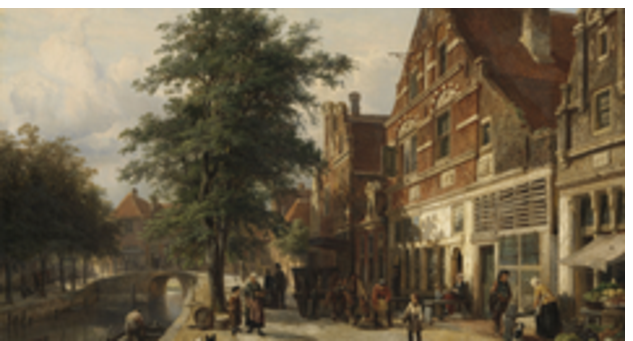}
\label{fig:label-B}}
\end{minipage}
\begin{minipage}{\linewidth}
\centering
\subfloat[Assembled image\newline(Conventional scheme,\newline
 $B_{x}=B_{y}=16$,\newline\!\!\!\!\!$D_{c}\!=\!0.25, N_{c}\!=\!0.26,
 L_{c}\!=\!0.24\!)$]{\includegraphics[height=2.2cm]{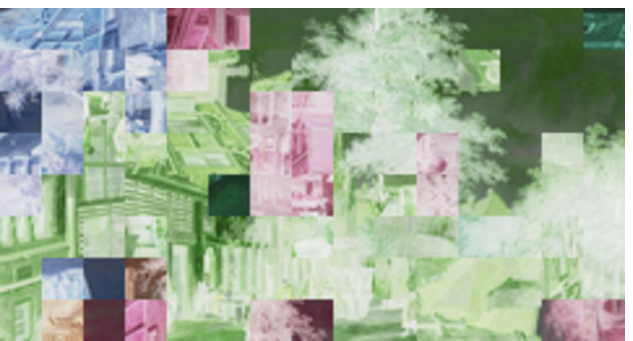}
 }
 \vspace{-0.1cm}
\centering
\subfloat[Assembled image\newline(Proposed
scheme,\newline\,\,$B_{x}=B_{y}=8$,\newline$D_{c}\!=\!0,N_{c}\!=\!0.002,\!L_{c}\!=\!0.002)$]{\includegraphics[height=2.2cm]{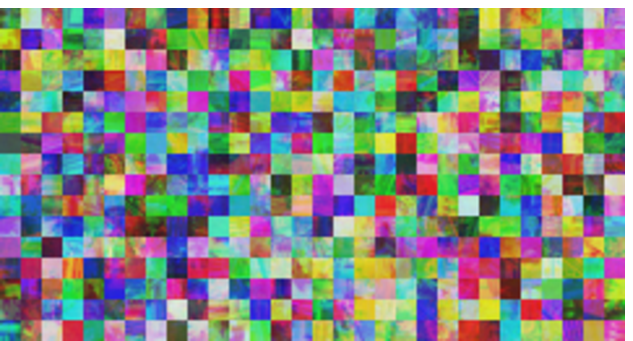}
\label{fig:label-C}}
\end{minipage}
\caption{Examples of assembled images}
\vspace{-0.6cm}
\label{fig:assembled}
\end{figure}

 \newcolumntype{M}{>{\centering\arraybackslash}p{2.1cm}}
  \newcolumntype{L}{>{\centering\arraybackslash}p{1.4cm}}
    \newcolumntype{K}{>{\centering\arraybackslash}p{1.5cm}}
\subsection{Results and discussions}
\label{res-diss}
The performances of the encryption scheme are shown in two aspects:
security, and quality of downloaded images.
\vspace{-0.35cm}\subsubsection{Security}
\vspace{-0.2cm}
\label{res-sec}
The security of the encryption schemes was evaluated by the robustness against
jigsaw puzzle solvers attack\cite{Gallagher_2012_CVPR,Son_2016_CVPR,Cho_2010_CVPR,CHUMAN2017ICASSP,CHUMAN2017ICME}.
This paper used the extended jigsaw puzzle
solver\cite{CHUMAN2017ICASSP,CHUMAN2017ICME} for assembling encrypted images.
There are three metrics \cite{Gallagher_2012_CVPR,Cho_2010_CVPR} using for determining the robustness against the extended solvers which are described as follows.
\begin{itemize}
	\item {\bf Direct comparison($D_c$)} is the ratio between the number of pieces which are placed in the correct position and the total number of pieces.
	\item {\bf Neighbor comparison ($N_c$)} expresses the ratio of the number of pieces that are joined with the correct pattern and the total number of pieces.
	\item {\bf Largest components($L_c$)} refers to the ratio between the number of
	the largest joined blocks that are correctly adjacent and the number of pieces.
\end{itemize}
In the measures, $D_c$, $N_c$, $L_c$ $\in [0,1]$, a larger value means a higher
compatibility as illustrated in Fig.\,\ref{fig:assembled}. Thirty different
encrypted images from dataset (b) were generated by random keys from one ordinary image.
The assembled image which has the highest sum of $D_c$, $N_c$, and $L_c$ in those of thirty images
was chosen. We performed these procedures for each original image independently,
and the average score of 20 images for each metric was calculated.
\par As shown in table\,\ref{table:jigsaw}, the average
$D_c$, $N_c$, and $L_c$ of images with the proposed scheme are respectively
equal to 0.002, 0.003, and 0.003 which are lower than those with the conventional
scheme. Summarily, encrypting images using the proposed scheme
provides more robustness against the extended jigsaw puzzle solver attack in
addition to larger key space.
 \begingroup
 \begin{table}[t]{
	\begin{center}
		\caption{Evaluation of the conventional and proposed scheme}
		\vspace{-0.3cm}
		\scalebox{1}{
		\begin{tabular}{|c|M|K|} \hline
			\multirow{2}{*}{Encryption types} & Conventional scheme\!\cite{kurihara2015encryption,KURIHARA2015} & Proposed scheme \rule[0mm]{0mm}{0mm} \\ \hline \hline
			Color channel & RGB & Gray scale \\ \hline
			Block size $B_{x}\times B_{y}$ & $16 \times 16$ & $8 \times 8$ \\ \hline
			Number of blocks $n$ & 120 & 1440 \rule[0mm]{0mm}{0mm}
			\\ \hline $D_{c}$ (Average) &0.189 &0.002  \rule[0mm]{0mm}{0mm}
			\\\hline $N_{c}$ (Average) &0.255 &0.003  \rule[0mm]{0mm}{0mm}
			\\\hline $L_{c}$ (Average) &0.310 &0.003  \rule[0mm]{0mm}{0mm} \\\hline
		\end{tabular}
		}
		\label{table:jigsaw}
	\end{center}
	\vspace{-0.8cm}
}
\end{table}
\endgroup
\begin{figure}[t]
\captionsetup[subfigure]{}
\centering
\subfloat[Twitter]{\includegraphics[clip, width=8cm]{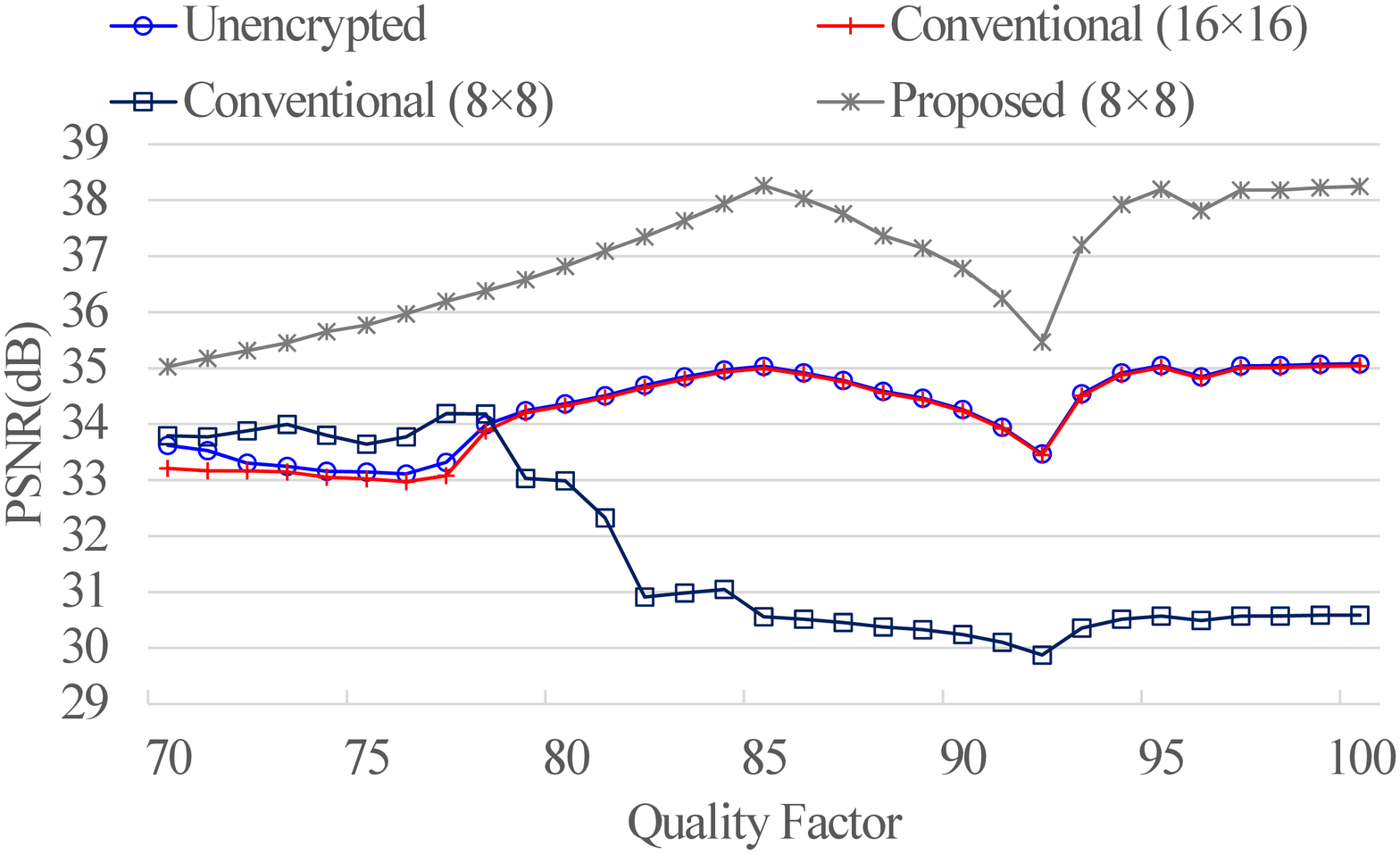}}
\vspace{-0.4cm}
\\
\subfloat[Facebook]{\includegraphics[clip, width=8cm]{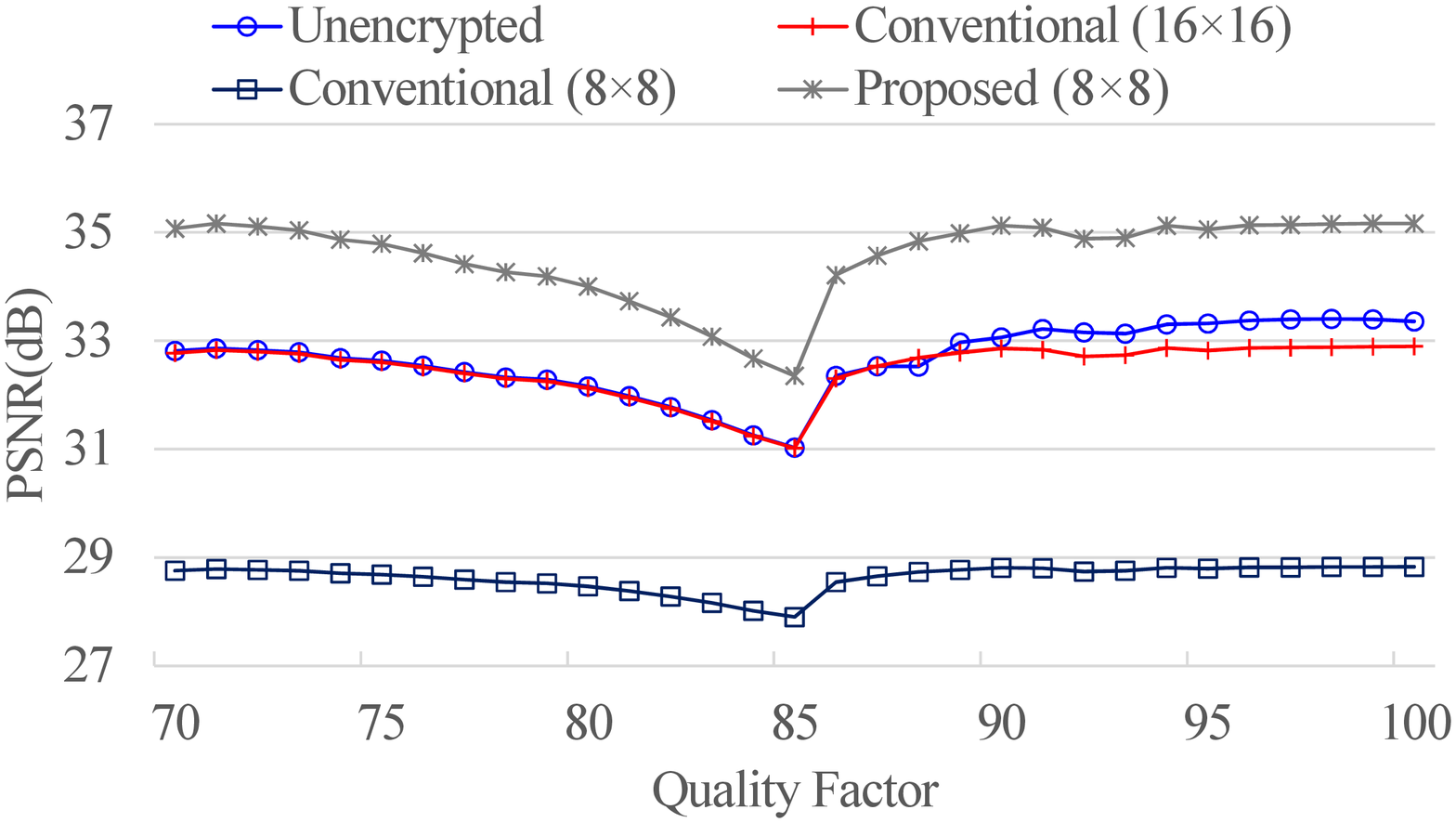}}
\vspace{-0.2cm}
\caption{The result of PSNR versus $Q_f$}
\label{fig:psnr-qf}
\vspace{-0.5cm}
\end{figure}
\vspace{-0.35cm}\subsubsection{Quality of downloaded images}
\vspace{-0.2cm}
\label{res-psnr}
To evaluate the effectiveness of the proposed scheme, we uploaded images
encrypted using the proposed scheme as well as the conventional
scheme\cite{kurihara2015encryption,KURIHARA2015} to Twitter and Facebook
respectively.
The images encrypted using the conventional scheme and the non-encrypted ones were
compressed with a 4:4:4 sub-sampling ratio and $Q_f$ =
70,71,\ldots,100. Then, downloading uploaded images from the SNS providers,
decoding the downloaded images, and decrypting the encrypted images were carried out respectively.\par Figure\,\ref{fig:psnr-qf} shows the performance for JPEG compressed
images without encryption, with the conventional scheme, and with the proposed
scheme. The arithmetic mean PSNR of 20 images from dataset (a) per quality
factor uploaded to Twitter and Facebook were plotted in
Fig.\,\ref{fig:psnr-qf}(a) and Fig.\,\ref{fig:psnr-qf}(b) respectively.
\par Twitter recompresses uploaded images that were compressed with high quality
factor ($Q_f$$\geqq$85), using $Q_f$ = 85 as shown in
table\,\ref{tb:sns_manipulation}\cite{CHUMAN2017APSIPA}. Otherwise, uploaded
JPEG images are not recompressed. In contrast, Facebook
recompresses every uploaded JPEG image regardless of quality factor based on its
compression algorithm with 4:2:0 sampling. \par As shown in
Fig.\,\ref{fig:psnr-qf}, the proposed scheme offered higher image quality than the conventional scheme even if image JPEG recompression was
carried out by Twitter and Facebook. When encrypting images using
conventional scheme with the same block size as the proposed scheme ($B_x = B_y = 8$), the
quality of downloaded images is strongly affected by the color sub-sampling
operation, while the proposed scheme can avoid this influence even in case of
using $B_x = B_y = 8$.
Moreover, the PSNR values of decrypted images with the proposed scheme were higher than unencrypted images in both providers due to avoiding the effect of
color sub-sampling. These results show that the proposed scheme allows us to
download higher quality images than unencrypted ones.

\section{Conclusion}
\label{conclusion}
This paper presented the new image encryption scheme that can be applied to SNS
called grayscale-based block scrambling image encryption. The proposed encryption
scheme can avoid the influence of color sub-sampling from JPEG compression.
Compared to the conventional scheme, the proposed scheme has less color
information, smaller block size, and larger number of blocks which enhance the
security and robustness against attacks. The experiment was conducted by
uploading the encrypted and unencrypted images to the SNS providers: Twitter and
Facebook. Moreover, the robustnesses against jigsaw puzzle solver attacks were
evaluated. The results proved that the proposed scheme offers the better
security and higher image quality than the conventional scheme.
\bibliographystyle{IEEEbib}
\begin{small}
\bibliography{refs}
\end{small}
\end{document}